# Levi-Civita Effect in the polarizable vacuum (PV) representation of general relativity

H. E. Puthoff

Institute for Advanced Studies at Austin
4030 W. Braker Ln., Suite 300, Austin, TX, 78759, USA
puthoff@earthtech.org

Claudio Maccone

Alenia Spazio S.p.A.
Via Martorelli 43, I-10155 Torino (TO), Italy

Eric W. Davis

Warp Drive Metrics
4849 San Rafael Ave., Las Vegas, NV, 89120, USA

**Abstract.** The polarizable vacuum (PV) representation of general relativity (GR), derived from a model by Dicke and related to the "$TH\varepsilon\mu$" formalism used in comparative studies of gravitational theories, provides for a compact derivation of the Levi-Civita Effect (both magnetic and electric), herein demonstrated.

## 1. INTRODUCTION

In the polarizable vacuum (PV) representation of GR, the vacuum is treated as a polarizable medium of variable refractive index [1]. Specifically, based on the isomorphism between Maxwell's equations in curved space and a medium with variable refractive index in flat space [2,3], the curved metric is treated in terms of variations in the permittivity and permeability constants of the vacuum, $\varepsilon_o$ and $\mu_o$, along the lines of the "$TH\varepsilon\mu$" formalism used in comparative studies of alternative gravitational theories [4]. The PV approach, introduced by Wilson [5], developed by Dicke [6,7], and recently elaborated by Puthoff [1], reproduces results predicted by GR for standard (weak-field) astrophysical conditions while posing testable modifications for strong-field conditions [1]. It is in application that the PV approach demonstrates its intuitive appeal and provides additional insight into what is meant by a curved metric. The derivation of the Levi-Civita Effect presented herein provides just such an example.

## 2. BACKGROUND

Paralleling the development of the standard tensor formulation of GR, alternative approaches have emerged that provide convenient methodologies for investigating metric changes in other formalisms. Of particular utility in calculating the magnitudes of GR effects while providing insight of an engineering nature is the formalism in which the dielectric constant of the vacuum plays the role of a variable refractive index under conditions in which vacuum polarizability alters in response to GR-type influences. In this approach the bending of a light ray near a massive body is seen as due to an induced spatial variation in the refractive index of the vacuum near the body, the reduction in the velocity of light in a gravitational potential is due to an effective increase in the refractive index, and so forth. This optical-engineering approach has been shown to be quite general [8,9], and to reproduce to required order both the equations of GR, and the match to the classical experimental tests of those equations.

## 3. METHODOLOGY

The PV methodology employed here follows that of Ref. [1]. As explained in detail there, the PV treatment of GR effects is based on an action principle (Lagrangian) that holds in special relativity, but with the modification that the velocity of light $c$ in the Lorentz factors and elsewhere is replaced by the velocity of light in a medium of variable refractive index, $c/K$; expressions such as $E = mc^2$ are still valid, but take into account that $c \rightarrow c/K$; and $E \left(= E_o/\sqrt{K}\right)$ and $m \left(= m_o K^{3/2}\right)$ are now functions of $K$; the vacuum polarization energy associated with the variable $K$ is explicitly included, and so forth. The Lagrangian density and resulting Euler-Lagrange particle and field equations that lead to GR-compatible results to testable order are given in Appendix A. For the Levi-Civita Effect, we examine Eq. (A-3) as it applies to empty space (static case),

$$\nabla^2 \sqrt{K} = -\frac{\sqrt{K}}{4\lambda}\left[\frac{1}{2}\left(\frac{B^2}{K\mu_o} + K\varepsilon_o E^2\right) - \frac{\lambda}{K^2}(\nabla K)^2\right]. \tag{1}$$

Here we see that changes in the vacuum dielectric constant $K$ are driven by the energy densities of the EM fields and the vacuum polarization.

## 4. LEVI-CIVITA EFFECT

Soon after Einstein published his theory of General Relativity, the Italian mathematical physicist Tullio Levi-Civita considered the possibility of the creation of an artificial gravitational field via generation of a static uniform magnetic or electric field (see Appendix B for a detailed description) [10,11]. In the context of the modern investigation of the theory of traversable wormholes [12] it was originally thought by one of us (Maccone) that the Levi-Civita spacetime metric would be considered a magnetic or electric field induced wormhole, and examined its implications for interstellar travel and communication [13,14]. However, it was later proved [15,16] that the Levi-Civita spacetime metric actually describes a spatial hypercylinder with a position dependent gravitational potential, and possessing none of the required characteristics of a traversable wormhole (see Appendix B). The geometry is nonetheless interesting from the standpoint that it describes a unique cylindrically shaped "trapped" space.

## 5. MAGNETIC HYPERCYLINDER SOLUTION IN THE PV FORMALISM

To examine the magnetic Levi-Civita Effect, we consider the case of a homogeneous, static magnetic field oriented in the $z$ direction (e.g., in the interior of a solenoid). Eq. (1) then takes the form

$$\frac{d^2\sqrt{K}}{dz^2} = \frac{1}{\sqrt{K}} \left[ \left( \frac{d\sqrt{K}}{dz} \right)^2 - \frac{4\pi G B^2}{\mu_0 c^4} \right] \quad , \tag{2}$$

where we have used $(\nabla K)^2 = 4K(\nabla \sqrt{K})^2$.

The solution to (2) takes the form

$$\sqrt{K} = \alpha \cos \beta z \quad , \tag{3}$$

where we have placed the maximum deviation of $K$ from unity at the origin, and the as-yet-undetermined integration constants $\alpha$ and $\beta$ are found to satisfy the constraint

$$\alpha^2 \beta^2 = \frac{4\pi G B^2}{\mu_0 c^4} \quad . \tag{4}$$

For even the strongest fields of interest (e.g., those of pulsars, $B \sim 10^9$ Tesla) perturbation of the spacetime metric is sufficiently small that Eq. (3) can be approximated by

$$\sqrt{K} \approx \alpha \left( 1 - \frac{\beta^2 z^2}{2} \right) \quad . \tag{5}$$

We now determine the constants $\alpha$ and $\beta$ by requiring that the velocity of light $c'(z) = c/K(z)$ transitions to $c$ (i.e., $K=1$) at a certain distance $L/2$ above and below the origin (e.g., at the ends of a solenoid of length $L$ centered at the origin). With the constants thereby determined, we obtain the solution

$$\sqrt{K} \approx 1 + \frac{2\pi G B^2}{\mu_0 c^4}\left[\left(\frac{L}{2}\right)^2 - z^2\right] \quad , \tag{6}$$

and thus the velocity of light $c'(z)$ within the magnetic field region is given by

$$\frac{c'(z)}{c} \approx \frac{1}{K} \approx 1 - \frac{4\pi G B^2}{\mu_0 c^4}\left[\left(\frac{L}{2}\right)^2 - z^2\right] \quad . \tag{7}$$

Therefore the velocity of light is slowed within the magnetic field, with its minimum value at the origin, equidistant from the ends of the magnetic field region. The transit time for a light ray through the magnetic field region is thus not $L/c$ but rather

$$\tau \approx \int_{-L/2}^{L/2} \frac{dz}{c'(z)} \approx \frac{L}{c}\left[1 + \frac{2\pi G B^2 L^2}{3\mu_0 c^4}\right] \quad . \tag{8}$$

## 6. ELECTRIC HYPERCYLINDER SOLUTION IN THE PV FORMALISM

The solution for the electric Levi-Civita Effect parallels the magnetic case. We consider a homogeneous static electric field oriented in the $z$ direction (e.g., between two charged plates). For the electric field case Eq. (1) takes the form

$$\frac{d^2\sqrt{K}}{dz^2} = \frac{1}{\sqrt{K}}\left[\left(\frac{d\sqrt{K}}{dz}\right)^2 - \frac{4\pi\varepsilon_0 G E^2}{c^4}\left(\sqrt{K}\right)^4\right] \quad . \tag{9}$$

The solution to (9) takes the form

$$\sqrt{K} = \alpha \operatorname{sech}\beta z \quad , \tag{10}$$

where again we have placed the maximum deviation of $K$ from unity at the origin. Here the integration constants $\alpha$ and $\beta$ satisfy the constraint

$$\beta^2 = \frac{4\pi\varepsilon_0 G E^2}{c^4}\alpha^2 \quad . \tag{11}$$

As in the magnetic case it can be assumed that for even the strongest fields of interest perturbation of the spacetime metric is sufficiently small that Eq. (10) can be approximated by

$$\sqrt{K} \approx \alpha\left(1 - \frac{\beta^2 z^2}{2}\right) \quad . \tag{12}$$

Following the procedure for the magnetic case, we require that the velocity of light transition from $c'(z) = c/K(z)$ to $c$ at $L/2$ above and below the origin (e.g., at charged plates spaced $L$ apart and centered at the origin). With the constants thereby determined, we obtain the solution for the electric hypercylinder case as

$$\sqrt{K} \approx 1 + \frac{2\pi\varepsilon_0 GE^2}{c^4}\left[\left(\frac{L}{2}\right)^2 - z^2\right] \quad . \tag{13}$$

The velocity of light $c'(z)$ within the electric field region is therefore given by

$$\frac{c'(z)}{c} \approx \frac{1}{K} \approx 1 - \frac{4\pi\varepsilon_0 GE^2}{c^4}\left[\left(\frac{L}{2}\right)^2 - z^2\right] \quad , \tag{14}$$

and the associated transit time for a light ray through the electric field region is

$$\tau \approx \int_{-L/2}^{L/2} \frac{dz}{c'(z)} \approx \frac{L}{c}\left[1 + \frac{2\pi\varepsilon_0 GE^2 L^2}{3c^4}\right] \quad . \tag{15}$$

## 7. CONCLUSION

In this short note we demonstrate how the polarizable vacuum (PV) formulation of general relativity (GR) lends itself to straightforward and compact derivations of GR effects, in this case the Levi-Civita Effect for both the magnetic and electric field induced hypercylinder space cases. Specifically, it is shown how the perturbation of the spacetime metric by the presence of uniform, static magnetic and electric fields can be understood as a perturbation of the effective refractive index of the vacuum, and summarized in terms of its effect on the propagation of a light ray through the region containing the fields.

## APPENDIX A

The Lagrangian density and resulting Euler-Lagrange particle and field equations that lead to GR-compatible results to testable order are given by Eqns. (32) – (34) of Ref. [1], viz:

$$L_d = -\left(\frac{m_o c^2}{\sqrt{K}}\sqrt{1-\left(\frac{v}{c/K}\right)^2} + q\Phi - q\mathbf{A}\bullet\mathbf{v}\right)\delta^3(\mathbf{r}-\bar{\mathbf{r}}) - \frac{1}{2}\left(\frac{B^2}{K\mu_o} - K\varepsilon_o E^2\right)$$
$$-\frac{\lambda}{K^2}\left[(\nabla K)^2 - \frac{1}{(c/K)^2}\left(\frac{\partial K}{\partial t}\right)^2\right],$$
(A-1)

$$\frac{d}{dt}\left[\frac{(m_o K^{3/2})\mathbf{v}}{\sqrt{1-\left(\frac{v}{c/K}\right)^2}}\right] = q(\mathbf{E}+\mathbf{v}\times\mathbf{B}) + \frac{(m_o c^2/\sqrt{K})}{\sqrt{1-\left(\frac{v}{c/K}\right)^2}}\left(\frac{1+\left(\frac{v}{c/K}\right)^2}{2}\right)\frac{\nabla K}{K}, \quad \text{(A-2)}$$

$$\nabla^2 \sqrt{K} - \frac{1}{(c/K)^2}\frac{\partial^2 \sqrt{K}}{\partial t^2} = -\frac{\sqrt{K}}{4\lambda}\left\{\frac{(m_o c^2/\sqrt{K})\left[1+\left(\frac{v}{c/K}\right)^2\right]}{\sqrt{1-\left(\frac{v}{c/K}\right)^2}}\delta^3(\mathbf{r}-\bar{\mathbf{r}})\right.$$

$$\left.+\frac{1}{2}\left(\frac{B^2}{K\mu_o}+K\varepsilon_o E^2\right)-\frac{\lambda}{K^2}\left[(\nabla K)^2+\frac{1}{(c/K)^2}\left(\frac{\partial K}{\partial t}\right)^2\right]\right\},$$

(A-3)

where $\lambda = c^4/32\pi G$, and $\varepsilon_o, \mu_o$ are expressed in MKSA units
.

We see in (A-2) that accompanying the usual Lorentz force is an additional dielectric force proportional to the gradient of the vacuum dielectric constant. This term is equally effective with regard to both charged and neutral particles and accounts for the familiar gravitational potential, whether Newtonian in form or taken to higher order to account for GR effects. In (A-3) we see that changes in the vacuum dielectric constant $K$ are driven by mass density (first term), EM energy density (second term), and the vacuum polarization energy density itself (third term). Eqns. (A-2) and (A-3), together with Maxwell's equations for propagation in a medium with variable dielectric constant, thus constitute the master equations to be used in discussing general matter-field interactions in a vacuum of variable dielectric constant as employed in the PV formulation of GR.

## APPENDIX B

Levi-Civita's spacetime metric for a static uniform magnetic field [10] was originally expressed by Pauli in the form [11]:

$$ds^2 = (dx^1)^2 + (dx^2)^2 + (dx^3)^2 + \frac{(x^1 dx^1 + x^2 dx^2)^2}{a^2 - \left[(x^1)^2 + (x^2)^2\right]}$$
$$-\left[c_1 \exp(x^3/a) + c_2 \exp(-x^3/a)\right]^2 (dx^4)^2$$

(B-1)

where $c_1$ and $c_2$ are integration constants that are determined by appropriate boundary conditions, and $x^1 \ldots x^4$ are Cartesian coordinates ($x^1 \ldots x^3$ = space, $x^4$ = time) with orthographic projection. The important parameter in (B-1) is:

$$a = \frac{c^2}{B\sqrt{4\pi G/\mu_o}} \approx 3.484 \times 10^{18}\frac{1}{B} \quad meters$$

(B-2)

which measures the radius of spacetime curvature induced by a homogeneous magnetic field with cylindrical symmetry (axis, $x^3 = z$) about the direction of the field (quantities expressed in MKSA units). Note that (B-1) is also the spacetime metric for a static uniform electric field in which the magnetic field intensity in (B-2) is replaced by the electric field intensity $E$ along with an appropriate change in the related vacuum electromagnetic constants (i.e., $B/\sqrt{\mu_o} \to E\sqrt{\varepsilon_o}$). The metric in (B-1) does not form a traversable wormhole according to [15,16]. To clarify this point we change the coordinate system from Cartesian to cylindrical coordinates $\left(x^1 = r\cos\varphi, x^2 = r\sin\varphi, x^3 = z, \text{let } x^4 = t\right)$, which transforms (B-1) into the form [15,16]:

$$ds^2 = -\left[c_1 \exp(z/a) + c_2 \exp(-z/a)\right]^2 dt^2 + \left(1 - \frac{r^2}{a^2}\right)^{-1} dr^2 + r^2 d\varphi^2 + dz^2 \tag{B-3}$$

This is a simpler form, and we can now inquire as to what spacetime geometry the Levi-Civita metric specifically describes. This is clarified this by a change of (radial) variable, $r = a\sin\theta$, $dr = a(\cos\theta)d\theta$, and substitution into (B-3) to obtain:

$$ds^2 = -\left[c_1 \exp(z/a) + c_2 \exp(-z/a)\right]^2 dt^2 + a^2\left[d\theta^2 + \sin^2\theta d\varphi^2\right] + dz^2 \tag{B-4}$$

where $a$ is the constant radius defined in (B-2). The spatial part of (B-4), $d\sigma^2 = a^2\left[d\theta^2 + \sin^2\theta d\varphi^2\right] + dz^2$, is recognized as the three-metric of a hypercylinder $S^2 \times \Re$. So (B-4) reveals that the Levi-Civita spacetime metric is simply a hypercylinder with a position dependent gravitational potential possessing no asymptotically flat region, no flared-out wormhole mouth and no wormhole throat.